\documentclass[preprint]{aastex}
\accepted{14 Oct 02 for Astron. J. }
\begin{document}
\title{ A Flaring L5 Dwarf: The Nature of H$\alpha$ Emission in 
Very Low Mass (Sub)Stellar Objects }

\author{James Liebert\altaffilmark{1}, J. Davy
Kirkpatrick\altaffilmark{2}, K. L. Cruz\altaffilmark{3}, I. 
Neill Reid\altaffilmark{4}, Adam Burgasser\altaffilmark{5},  
C. G. Tinney\altaffilmark{6}, and John E. Gizis\altaffilmark{7}}

\altaffiltext{1}{Steward Observatory, University of Arizona, 
Tucson, AZ  85721, liebert@as.arizona.edu}

\altaffiltext{2}{Infrared Processing and Analysis Center, California
Institute of Technology, Pasadena, CA 91125, davy@ipac.caltech.edu}

\altaffiltext{3}{Dept. of Physics \& Astronomy, Univ. of Pennsylvania,
209 South 33rd St., Philadelphia, PA 19104, kelle@sas.upenn.edu }

\altaffiltext{4}{Space Telescope Science Institute, 
3700 San Martin Dr., Baltimore MD 21218; and Department of 
Physics and Astronomy, Univ. of Pennsylvania, 209 S. 33rd St., 
Philadelphia, PA 19104-6396, inr@stsci.edu}

\altaffiltext{5}{UCLA, Division of Astronomy and Astrophysics, 
8965 Mathematical Sciences Bldg., Los Angeles, CA 90095-1562, 
adam@astro.ucla.edu}

\altaffiltext{6}{Anglo-Australian Observatory, P.O. Box 296, 
Epping, 1710, Australia, cgt@aaoepp.aao.gov.au}

\altaffiltext{7}{Department of Physics and Astronomy, Univ. of 
Delaware, Newark, DE 19716, Gizis@udel.edu}

\begin{abstract}

Times series spectrophotometry of the L5 dwarf 2MASS~01443536-0716142
showed strong H$\alpha$ emission which declined by nearly 75\% in four
consecutive exposures.  The line was not detected in emission on a
spectrum obtained eleven months later.  This behavior contrasts with
that of 2MASSI~J1315309-264951, an L5 dwarf which has shown even
stronger H$\alpha$ emission on four separate occasions.  The
observational database suggests that L dwarfs can be found in such
strong flares only occasionally, with a duty cycle of order 1\%.  In
contrast, the few, continuously-strong H$\alpha$ emitters, including
PC~0025+0447 and 2MASSI~J1237392+652615, must either be (1) objects no
older than 10-100~Myrs with continuously-active accretion and/or
chromospheres, but which apparently formed in isolation from known young
stellar clusters and associations, or (2) objects empowered by a
different but unknown mechanism for the H$\alpha$ energy.

\end{abstract}


\section{Introduction }

With the recent discoveries of numerous late M, L and T dwarfs, it has
become apparent that the level of chromospheric activity as measured by
the H$\alpha$ emission line flux relative to the total bolometric flux
generally declines rapidly later than about M6-7 spectral type (Gizis et
al. 2000, Basri 2000). In the former reference, H$\alpha$ emission was
not even detected in any dwarf from the 2MASS sample later than L3.
Similar trends have been apparent in the DENIS and SDSS discoveries of L
dwarfs (Delfosse et al. 1997, Phan-Bao et al. 2001, Hawley et al. 2002).
Only two of the few dozen 2MASS T dwarfs (Burgasser et al. 2002b) showed
a detected emission line in low resolution spectra, where admittedly
most of these are single-epoch observations.  Upper limits for the
H$\alpha$ emission of late L and T dwarfs relative to the estimated
bolometric luminosities are typically an order of magnitude below the
detected fluxes of late M dwarfs.  The simple interpretation is that
late L and T dwarfs generally exhibit weak or no chromospheric activity.
The T dwarf 2MASSI~J1237392+652615 (hereafter 2M1237) is an
extraordinary exception to this trend: it has shown strong H$\alpha$
emission on each of several nights it has been observed (Burgasser et
al. 2000,2002a), in a manner similar to the M9.5 dwarf PC~0025+0447
(Schneider et al. 1991; Mart\'{\i}n, Basri \& Zapatero Osorio 1999,
hereafter PC0025).

Another property of the chromospherically-active late M dwarfs is the
occurrence of flares, observed at ultraviolet, X-ray and radio
frequencies as well as via optical spectra or photometry.  Bursts of
X-ray emission interpreted as coronal flares were detected in Chandra
observations of the long-known M8V dwarf VB~10 (Fleming, Giampapa, \&
Schmitt 2000) and the young M9 brown dwarf LP~944-20 (Rutledge et
al. 2000).  The former exhibited a far-ultraviolet flare observed by
Linsky et al. (1995) using the GHRS on the Hubble Space Telescope.
Likewise LP~944-20 has been observed to emit bursts at radio
frequencies (Berger et al. 2001).

A 2MASS M9.5 dwarf (2MASSW~J0149090+295613, hereafter 2M0149)
exhibited very strong H$\alpha$ and excess red continuum emission for
a few tens of minutes, but then the emission line weakened and the
normal photospheric absorption spectrum was observed (Liebert et
al. 1999).  If larger amounts of energy were radiated at far-UV and
X-ray wavelengths as found in multi-wavelength studies of M flare
stars (cf. Eason et al. 1992), the total energy release could have
approached or exceeded the normal photospheric luminosity for $\sim$10
minutes.  This object has subsequently been observed several times in
the course of spectrophotometric surveying of 2MASS sources, and has
exhibited only weak H$\alpha$ emission.

Another M9.5 dwarf, BRI~0021-0214 (hereafter BRI0021), was observed
to flare by Reid et al. (1999).  This rapidly-rotating object
($v_{sin~i}~\sim$40 km~s$^{-1}$) generally shows no H$\alpha$ emission
(Basri \& Marcy 1995).  This underscores the probability that many of
the latest M and perhaps L dwarfs exhibiting no apparent H$\alpha$
emission on their classification spectra may at times show strong
emission in flares.

Hall (2002a) found the first L dwarf to show strong H$\alpha$ emission,
2MASSI~J1315309-264951 (hereafter, 2M1315), which he estimated to be of
type L3.  This emission was also present though a factor of two weaker
at a second observation five months later.  It was thus not clear on
what time scale the variability takes place, but Hall argued that
chromospheric activity was the likely cause, and that this may be the
first observation of a flaring L dwarf.  Gizis (2002) independently
found this object to show strong H$\alpha$ emission for six contiguous
600 second exposures on one night and two consecutive 900 second
exposures on the following night.  Since the emission was constant over
those time periods, Gizis argued that the object is an L analog to the
anomalous T emission object discussed in the first paragraph.  Gizis
classified the dwarf as L5 (and we adopt this subtype here).  Finally, 
Hall (2002b) reports a fourth epoch of spectroscopy, again finding 
very strong H$\alpha$ emission. 

In this paper we present data for the L5 dwarf 2MASS~01443536-0716142
(hereafter, 2M0144) which we show to be the first L-dwarf detected in an
unambiguous flare.  In section 2 we present the time series of
spectrophotometric data.  Section 3 is a brief discussion of the
chromospheric properties and variability of the relatively few known
active L dwarfs.

\section{Observations}

2M0144 is one of the brightest of the 2MASS L dwarfs with J=
14.19$\pm$0.04, H= 13.01$\pm$0.04, and K$_s$= 12.27$\pm$0.04.  Its
J-K$_s$ = 1.92$\pm$0.06 is consistent with the spectral type.

2M0144 was first observed spectroscopically on 20 Feb 2001 with the Low
Resolution Imaging Spectrophotometer (LRIS, Oke et al. 1995) on the
Keck~I telescope.  These spectra were reduced using the IRAF package and
techniques described in Kirkpatrick et al. (1999). A second, separate
observation was obtained with the CTIO 4m Blanco reflector on 24 Jan
2002; the IRAF data reduction techniques used were similar to the
description in Cruz \& Reid (2002).  Because of an error in the flux
calibration at the red end of the CTIO spectrum, however, an empirical
adjustment was derived from a comparison of the Keck and CTIO spectra to
correct the flux scaling of the CTIO data.  In adopting this procedure,
we make the assumption that the long-wavelength continuum of the
spectrum is not variable.  
 
These spectra are shown in Figure~1.  The Keck spectra were a
consecutive sequence of two 180 second exposures followed by two more
300 second integrations.  The CTIO spectrum was a single 1800 second
exposure.  It is readily seen that the H$\alpha$ is strongest in the
first scan, but is nearly 75\% weaker on the last integration of the
Keck dataset.  Only an upper limit was measured from the CTIO spectrum.
The NOAO Image Reduction and Analysis Facility ($IRAF$) ``splot''
routine was used to estimate equivalent widths and line fluxes. For each
UT starting time, these measurements are listed in Table~1, with the
fluxes in units of $10^{-16}$ erg cm$^{-2}$ s$^{-1}$.

The measurements of the H$\alpha$ and the continuum fluxes depend 
on both the numbers of photons detected and the accuracy of the 
absolute calibration of the spectrophotometry.  Since the continuum 
may be measured over a fairly wide bandpass about the line position, 
the photon flux accuracy is certainly a few per cent, as is the line 
flux at the peak of the flare.  However, the accuracy of the 
spectrophotometry is limited by the use of a mean extinction curve, 
and especially by the ability to center the faint star on the slit 
using the acquisition TV.  We therefore estimate that the peak 
H$\alpha$ and continuum fluxes are measured to 10\% accuracy for 
2M0144, the main subject of this paper.  However, for a really 
faint target like the T dwarf 2M1237, these measurements may be 
no better than 30\%.   

2M0144 clearly exhibits the characteristics of flaring, with a time
scale similar to that shown on the data set for 2M0149 (Liebert
et al. 1999).  However, the flare was much weaker in total amplitude
than exhibited by the M9.5 dwarf.  A He~I 6678\AA\ emission line was
measured at 3.5$\pm$2\AA\ EW or 1.5$\times$10$^{-17}$ erg cm$^{-2}$
s$^{-1}$ on the first Keck exposure, but the spectrum is too noisy for
this to be a definite detection.  The short wavelength end of the first
spectrum exhibited no continuum enhancement relative to the ones with
weaker H$\alpha$ emission.
\begin{deluxetable}{lcc}

\tablenum{1}
\tablecaption{ Individual H$\alpha$~Strengths }
\tablewidth{0pt}
\tablehead{
\colhead{Epoch} & \colhead{EW(\AA)} & \colhead{ f$_{H_\alpha}$ }  }

\startdata

20~Feb~01~05:14 & 23 & 2.4 \\
20~Feb~01~05:20 & 16 & 1.8 \\
20~Feb~01~05:24 & 12 & 1.3 \\
20~Feb~01~05:30 &  6 & 0.8 \\
24~Jan~02~01:00 & $<$3 & $<$0.4 \\

\enddata
\end{deluxetable}
 
Finally, we coadded the four Keck~I spectra to maximize S/N near the
Li~I 6707\AA\ resonance doublet.  We estimate an upper limit equivalent
width of 2\AA\ for this absorption feature.  In Kirkpatrick et
al. (2000), about 40\% of the well-observed dwarfs near type L5 show
easily detectable lithium (at $\sim$10\AA), an unambiguous signature of
an interior incapable of hydrogen fusion.  Thus, we cannot claim that this
is a brown dwarf, although it may well be a massive brown dwarf capable
of depleting lithium but not burning hydrogen.  Note that the
published spectra taken of 2M1315 probably measure the continuum too
weakly at these wavelengths for a strong lithium absorption feature to
be detected.

The 9400\AA\ H$_2$O band absorption appears substantially stronger on
the CTIO spectrum than on the Keck scans.  Since no attempt was made to
remove telluric water absorption on either observation, we cannot rule
out a terrestrial origin for this apparent variability.  

\section{Discussion}

As mentioned in the Introduction, it is customary to consider the ratio
of the H$\alpha$ flux to the estimated bolometric flux
(f$_{H\alpha}$/f$_{bol}$) as a measure of the relative strength of this
activity.  Since the spectral type of 2M0144 is L5, we adopt for the
bolometric correction to the K magnitude (BC$_K$) the value of 3.3
magnitudes determined for the well-observed prototype L4 dwarf GD~165B.
This quantity varies only slowly with subtype through the L sequence
(Leggett et al 2001, Reid et al. 2001).  We thus estimate an apparent
bolometric magnitude of m$_{bol}$= 16.0, yielding
f$_{H\alpha}$/f$_{bol}$ in the range 0.8--2.4$\times$10$^{-5}$ for the
20~Feb~01 flare sequence.  The upper limit from the CTIO spectrum
corresponds to a flux ratio of 0.4$\times$10$^{-5}$.  We estimate that
the bolometric flux of 2M0144 is estimated to 5\% accuracy, so that the
ratios may be measured to better than 15\%.

In Figure~2 we plot the log of this ratio vs. spectral type for several
H$\alpha$-active objects discussed here, in comparison with late M
dwarfs (X's and upper limits), L dwarfs (open circles and upper limits)
and T dwarfs (open square and upper limits) from the data sets of
Hawley, Gizis and Reid (1996), Cruz and Reid (2002), Gizis et al. (2000)
and Burgasser et al. (2002b).  Here a numerical code is used for the
spectral type where 0, 10 and 20 correspond to M0, L0 and T0, as shown.
The errors in the measurements of f$_{H\alpha}$ and f$_{bol}$, which add
uncertainty to the range exhibited of the ratio, are not displayed in
the figure.

The full range from maximum to quiescence is shown by the connected
points for the three flare stars observed in time series, 2M0149,
BRI0021 and 2M0144 (filled squares); in quiescence, these show fluxes
typical of other late M and L dwarfs.  Note that virtually all of the
fluxes shown for the other M, L and T dwarfs shown here were obtained in
quiescence, yet one wonders if many/all of them can exhibit similar
flares.  In contrast, the maximum and minimum of four epochs of
observation of 2M1315 (asterisks) by Hall (2002a) and Gizis (2002) all
lie in a range two orders of magnitude higher than the limits for all
but a few middle/late L dwarfs.  Its discovery likely adds a third field 
object to the possible class (with PC0025 and 2M1237, also plotted as
asterisks) exhibiting a continual or persistent state of high H$\alpha$
emission (Gizis 2002).  The factor of four range in H$\alpha$ flux for
PC0025 reported by Mart\'{\i}n, Basri, \& Zapatero Osorio (1999) is shown,
and the BC$_K$ = 3.19 for LHS~2924 (M9V) obtained by Tinney, Mould, \&
Reid (1993) is used to compute the bolometric magnitude.  It is not
clear that the small range observed from fewer measures of the H$\alpha$
flux of 2M1237, given the larger uncertainty in its H$\alpha$ flux 
mentioned earlier, represents real variability (Burgasser et al. 2002a).  

In contrast to the objects exhibiting continually strong emission,
2M0144 was observed with an f$_{H\alpha}$ flux more than a factor of
10 higher than the mean (including mostly upper limits) for L dwarfs
of similar spectral type.  It may therefore be the only true H$\alpha$
flare star yet found later than M9.5.  Yet, at least 100 spectra of L
dwarfs have been taken of 2MASS candidates alone (Kirkpatrick et
al. 2000), another 48 are reported in Hawley et al. (2002), and at
least a few dozen in DENIS papers in order to discover and classify
these objects.  That virtually all of these were caught in a quiescent
state indicates that magnetic eruptions are indeed rare, ie. occur
only of the order $\sim$1\% of the time.  This ``duty cycle'' estimate
applies only to the fraction of the time spent in strong emission
(ie. $\ga$ an order of magnitude stronger than the mean).  Using a
similar criterion, Gizis et al. (2000) estimated that very late M
dwarfs spend $\ga$7\% of the time in a flare state, very similar to
the duty cycle for BRI0021 alone estimated by Reid et al. (1999) from
$\sim$36,600 seconds of published observations.

Mohanty et al. (2002) suggest that the very high electrical
resistivities in the cool, predominantly-neutral atmospheres of L dwarfs
cause the drop-off in chromospheric activity beginning with late M
dwarfs.  Earlier a similar argument had been made to account for the
weak, quiescent coronal X-ray fluxes of late M dwarfs by Fleming,
Giampapa, \& Schmitt (2000). The increasingly-cool boundaries may
inhibit the magnetic field flux tubes from penetrating the surface to
cause flares.  This might account for the apparently decreasing flare
``duty cycle'' with decreasing T$_{eff}$.  The relationship of the
recently-discovered outbursts of radio emission in cool dwarfs as late
as L3.5 (Berger 2002) to the H$\alpha$ flares remains unclear.

The bigger puzzle may be in explaining the few field objects showing
persistent, strong H$\alpha$ emission.  Statistically, these seem also
to be of the order of 1\% of the known samples of very late M, L and T
dwarfs.  

First, we consider the hypothesis that continuous flaring or
chromospheric/coronal energy release by an extremely active magnetic
dynamo is taking place, as Mart\'{\i}n, Basri, \& Zapatero Osorio (1999)
argued for PC0025.  However, these authors measured $v~sin~i$ of only 13
km s$^{-1}$ for PC0025.  This is considerably less than the $\sim$40 km
s$^{-1}$ measured for relatively inactive objects like BRI0021.  Thus,
the nature of the dynamo may differ from the rotation-driven ``shell''
dynamo believed to be operative in the Sun and late-type dwarfs with
radiative cores.  Durney, De Young \& Roxburgh (1993) proposed a
turbulent (or distributed) dynamo model which could operate in
completely-convective interiors while having a weaker effect on the
internal angular momentum.  It is not clear, however, why a ``special''
dynamo would operate in just a few of the ultracool dwarfs, all of 
which are believed to have completely-convective interiors.

Alternatively, we have suggested that the emission in these objects is
caused by their being interacting binaries (Burgasser et al. 2000).
However neither Mart\'{\i}n, Basri, \& Zapatero Osorio (1999, for PC0025)
nor Burgasser et al. (2002a, for 2M1237) found any evidence in favor
of the binary hypothesis.

Finally, we consider the possibility that the strong H$\alpha$
emitters are very young objects.  Objects with M7-9 spectra in very
young associations and clusters are observed to show such strong
emission -- such as 162349.8-242601 in $\rho$~Oph (Luhman, Liebert, \&
Rieke 1997), CFHT-BD-Tau J043947.3+260139 in Taurus (Mart\'{\i}n et
al. 2001), S~Ori~55 and 71 in the Sigma Orionis cluster (Zapatero
Osorio et al. 2002a, Barrado y Navascu\'es et al. 2002), and probably
2MASSW J1207334-393254 in TW~Hya (Gizis 2002).  We plot the strongest
of these sources, S~Ori~71, as an asterisk in Figure~2.  These all
belong to stellar systems which are arguably no older than 10~Myr.
Accretion may be the dominant contributor to the emission in these
objects, although likely driven by a strong protostellar magnetic
field (Shu et al. 1994).  

Gizis (2002) found 2M1315 because it was in the wide area of the
TW~Hya association he was surveying.  Unlike the aforementioned
objects, however, he found that the spectrum did not show photospheric
absorption features indicative of a low gravity, and therefore the
object is not likely to be a member of the association.

There is no such ambiguity with PC0025 and 2M1237: these objects do not
lie in or near any apparent site of recent star formation.  Moreover, if
one assumed an age of only 10~Myr, the models of Burrows et al. (2002,
see Fig.~8) suggest that PC0025 should be barely massive enough to fuse
deuterium (15-20~M$_J$).  For 2M1237 to have faded to a likely T$_{eff}$
below 1,000~K within 10~Myr, the models suggest it would have a mass of
3-4M$_J$!  

Spectroscopic studies of brown dwarf candidates in clusters an order of
magnitude older -- Alpha Per at $\sim$90~Myr (Stauffer et al. 1999), and
the Pleiades at $\sim$120~Myr (Stauffer, Schultz, \& Kirkpatrick 1998)
-- have shown that these clusters do not harbor objects of late M
spectral class exhibiting H$\alpha$ emission stronger than about 10\AA\
equivalent width.  A similar conclusion was reached by Gizis, Reid, \&
Hawley (2002) for nearby field dwarfs: by an approximate age of 10$^8$
years, though less rigorously estimated than for a cluster, the samples 
of M dwarfs do not include objects with H$\alpha$ equivalent widths
$>$10\AA.
 
There is nonetheless an order of magnitude difference in age between
Rho Oph, Taurus, Sigma Ori, and TW~Hya ($<$10~Myr) and Alpha
Per/Pleiades ($\sim$100~Myr).  It appears that examples of extreme
H$\alpha$ emission, which exist at 10~Myr, disappear by 100~Myr.  But
at what intermediate age does the strong emission fade?  A few
clusters (IC~2602, IC~2391) with 30-50~Myr ages have been surveyed
photometrically, and large numbers of substellar candidates identified
(Barrado y Navascu\'es et al. 2001a), but spectroscopic investigation
of the low mass potential members has apparently not yet shown whether
extreme, persistent H$\alpha$ emitters are present.

The most straightforward, but perhaps naive, conclusion may be that
PC0025, 2M1237 and probably 2M1315 are rare objects of ages 10-100~Myr
in the solar neighborhood.  If the T6 dwarf 2M1237 formed 100~Myr ago,
the Burrows et al. (2001) tracks suggest a mass of $\sim$12~M$_J$,
marginal for a deuterium-burning phase.  The apparent discovery of such
low mass objects in the Sigma Orionis cluster (Barrado y Navascu\'es et
al. 2001b, Zapatero Osorio et al. 2002b) suggests that they can form in
isolation, at least within a cluster.  However, the lack of any apparent
association with a known, young stellar association or cluster leaves us
uncomfortable with this conclusion.

In summary, the great majority of L dwarfs exhibit much less
chromospheric activity than their more massive counterparts.  They can
exhibit flares, but the duty cycle appears to be only of order 1\%.
Three of the known objects showing persistent, strong H$\alpha$
emission are either young objects which managed to form in relative
isolation from known young associations and clusters; or, perhaps more
likely, may be exhibiting a different, as-yet unknown mechanism for
the line emission.

\acknowledgements

We thank the referee, Suzanne Hawley, for several helpful suggestions
and comments, and acknowledge insightful discussions with Mark Giampapa.
This publication makes use of data products from the Two Micron All Sky
Survey, which is a joint project of the University of Massachusetts and
the Infrared Processing and Analysis Center/California Institute of
Technology, funded by the National Aeronautics and Space Administration
and the National Science Foundation.  The data presented herein were
obtained at the W.M. Keck Observatory, which is operated as a scientific
partnership among the California Institute of Technology, the University
of California, and NASA. The authors also wish to recognize and
acknowledge the very significant cultural role and reverence that the
summit of Mauna Kea has always had within the indigenous Hawaiian
community.  We are most fortunate to have had the opportunity to conduct
observations from this mountain.  The NStars research described in this
paper was supported partially by a grant awarded as part of the NASA
Space Interferometry Mission Science Program, administered by the Jet
Propulsion Laboratory, Pasadena.  AJB acknowledges support by NASA
through Hubble Fellowship grant HST-HF-01137.1 awarded by the Space
Telescope Science Institute, which is operated by the Association of
Universities for Research in Astronomy, Inc., for NASA, under contract
NAS 5-26555.  KCL acknowledges support from an NSF Graduate Research
Fellowship.  This research is supported by a NASA JPL grant (961040NSF).

Finally, as we write this on the eve of the final release of the all-sky
2MASS data, we thank in particular the survey Principal Investigator
Mike Skrutskie, Project Manager Rae Stiening, and the NASA IPAC staff
led by Roc Cutri, for all their efforts that made studies like this
possible.  We also add particular thanks to Susan Kleinmann and Frank
Low, without whose vision and energy the project might never have gotten
started.

\newpage

\begin{figure} 

\caption{Four Keck~I LRIS spectra of 2MASS~01443536-0716142 on 20 Feb
2001 (from top), with starting UT times labelled.  The decline in
H$\alpha$ at 6563\AA\ is evident, with fluxes given in Table~1.  The
bottom spectrum obtained eleven months later at the CTIO Blanco 4~m
shows no detected emission line. }

\end{figure} 

\begin{figure}

\caption{The logarithmic ratio of f(H$\alpha$) to f(bol) is plotted
against spectral type for several unusual flaring and emission line
objects of types M9.5 to T6.  Shown for comparison are corresponding
values including upper limits from samples of M, L and T dwarfs.
Different symbols shown here are discussed in the text.  The three
objects considered to undergo flares are plotted as filled squares.  The
lower values for BRI0021 and 2M0144 are upper limits.  The horizontal
positions of the 2M0149 filled squares have been adjusted slightly to
avoid confusion. Three objects exhibiting strong emission on all spectra
are plotted as asterisks. }

\end{figure}

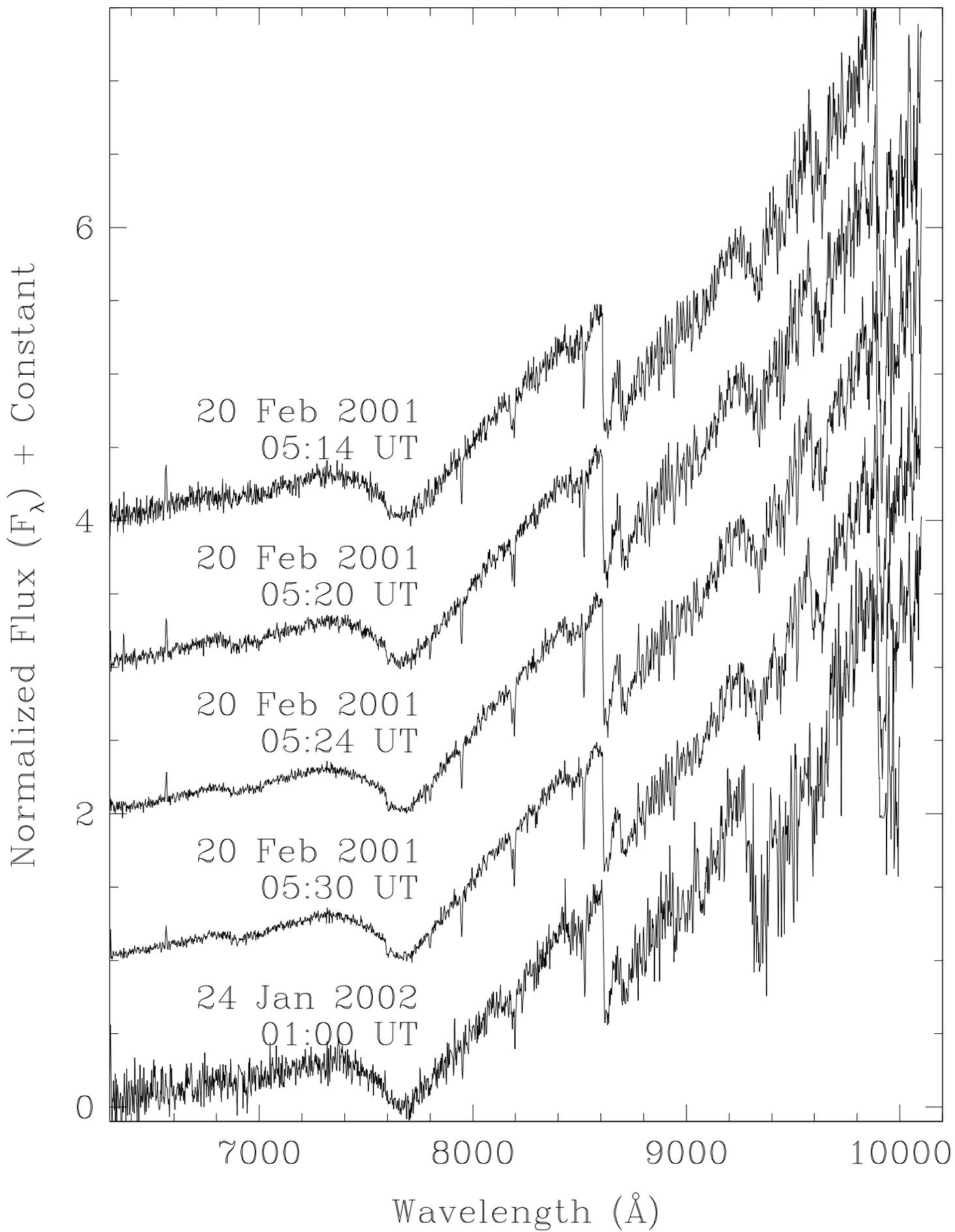
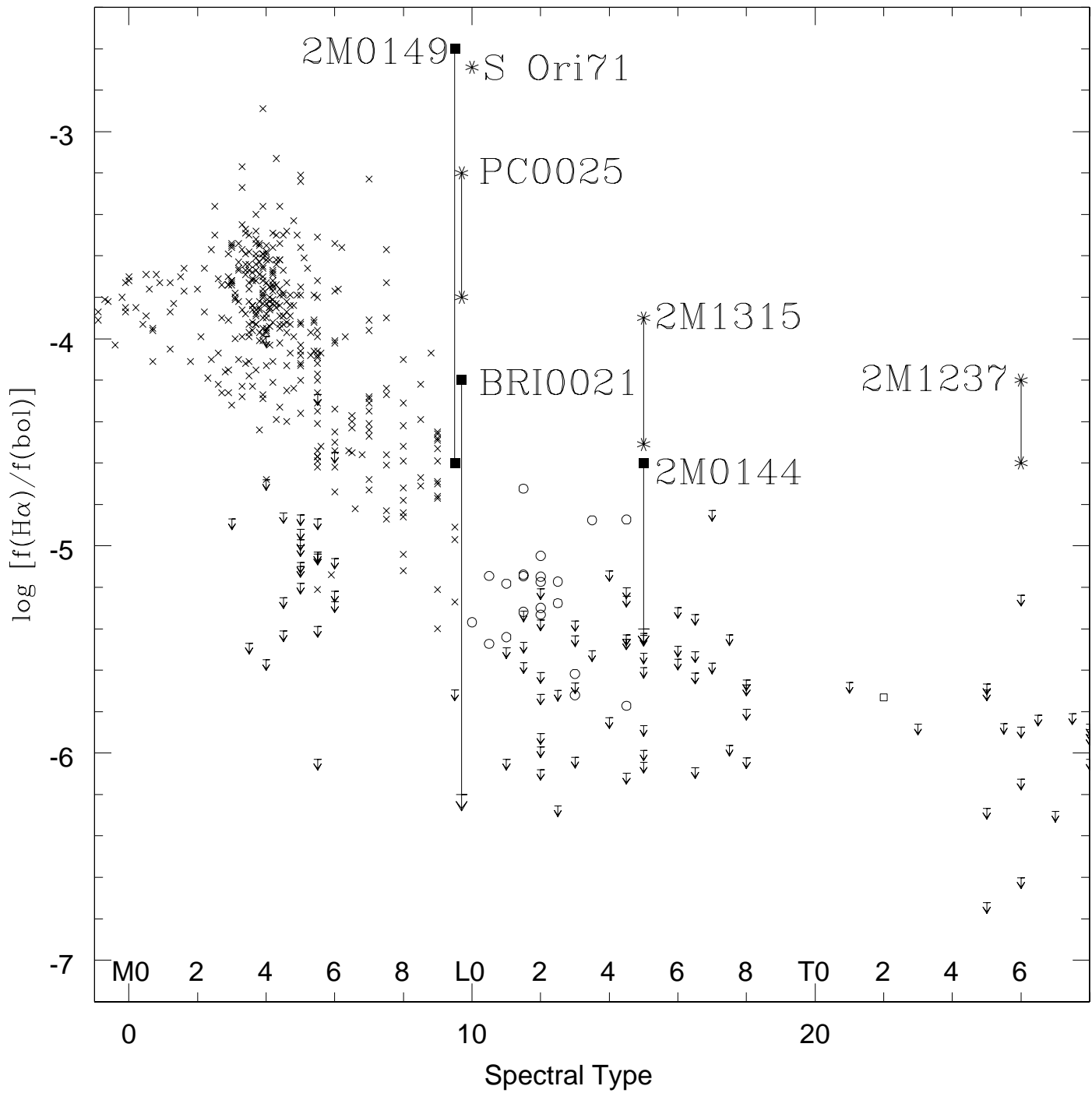
\end{document}